# Vector Quantized Spectral Clustering applied to Soybean Whole Genome Sequences

Aditya A. Shastri[1], Kapil Ahuja[1], Milind B. Ratnaparkhe[2], Aditya Shah[1], Aishwary Gagrani[1], Anant Lal[1]

[1]Computer Science and Engineering, Indian Institute of Technology Indore, Khandwa Road, Simrol, Indore, Madhya Pradesh 453552, India

[2]ICAR-Indian Institute of Soybean Research, Khandwa Road, Indore, Madhya Pradesh 452001, India

**Abstract:** We develop a Vector Quantized Spectral Clustering (VQSC) algorithm that is a combination of Spectral Clustering (SC) and Vector Quantization (VQ) sampling for grouping Soybean genomes. The inspiration here is to use SC for its accuracy and VQ to make the algorithm computationally cheap (the complexity of SC is cubic in-terms of the input size). Although the combination of SC and VQ is not new, the novelty of our work is in developing the crucial similarity matrix in SC as well as use of $k$-medoids in VQ, both adapted for the Soybean genome data. We compare our approach with commonly used techniques like UPGMA (Un-weighted Pair Graph Method with Arithmetic Mean) and NJ (Neighbour Joining). Experimental results show that our approach outperforms both these techniques significantly in terms of cluster quality (up to 25% better cluster quality) and time complexity (order of magnitude faster).

# Introduction

Clustering is one of the most widely used techniques for data analysis having applications in almost every field like statistics, computer science, biology, social sciences, psychology, etc. People attempt to get a first impression of their data by trying to identify groups having similar behaviour. Finding tight clusters, i.e., well separated and compact, is very important. Commonly used clustering algorithms include $k$-means, PAM, CLARA, BIRCH, DBSCAN, Wave-Cluster, EM, etc[1]. Compared to these traditional algorithms, a promising alternative is to use spectral methods for clustering.

Clustering algorithms that use spectral properties are widely used because of their accuracy (we get more tight clusters) and easy implementation (these algorithms can be solved efficiently by using standard linear algebra methods)[2]. However, when the input data is very large, they become inefficient; computational complexity of $O(n^3)$, where $n$ is the size of the input data. Hence, considerable research has been done to reduce this complexity without affecting the accuracy of the underlying algorithm.

One such method is sampling that can reduce the input size. Samples should be selected in a manner such that they represent the whole dataset uniformly. Many techniques exist for sampling like random sampling, stratified sampling, matrix factorization, vector quantization, pivotal sampling, the strip method, the mean method, the second derivative method etc[3, 4]. Among these, vector quantization (VQ)[5] is commonly used and is easy to implement since it provides the reduced data in a single scan of elements.

Clustering of Whole Genome Sequences (WGSs[*]) is useful in developing better species of plants e.g., disease resistant, drought resistant etc. Here, the traditional methods for clustering, e.g., Un-weighted Pair Graph Method with Arithmetic Mean (UPGMA)[6] and Neighbour-Joining (NJ)[6], which are currently used by plant biologists, do not provide the level of accuracy needed and are also not the most efficient methods because of their high computational complexity ($O(n^3)$).

In this paper, we use the Spectral Clustering (SC) algorithm (for accuracy) along with VQ (for efficiency) for clustering Single Nucleotide Polymorphism (SNP[†]) data obtained from the WGSs of Soybean plant. Although this combination of SC and VQ is not new, the novelty of our work is using the two for clustering SNP data.

One very important step in the SC algorithm is construction of the matrix called the similarity matrix. We implement our algorithm by constructing four different similarity matrices for Soybean SNP data. The original VQ uses $k$-means to sample the data. However, our data is in the form of strings of A, T, G, C and the mean for this does not exist. Hence, we modify VQ and use $k$-medoids to sample sequences.

We test our algorithm on SNP sequences obtained from a standard Soybean database. We compare our results with currently used methods of clustering SNP data (mentioned above). For a fair comparison in-terms of efficiency (due to sampling) of all algorithms involved, we develop vector quantized versions of these existing methods as well, i.e.

[*] WGS is a sequence made from a combination of 4 nucleotides; A (Adenine), T (Thymine), G (Guanine), and C (Cytosine)[7].

[†] SNP is the variation in the nucleotide that occurs at a specific position across sequences.

Vector Quantized UPGMA (VQUPGMA) and Vector Quantized NJ (VQNJ). Experiments show that our Vector Quantized Spectral Clustering (VQSC) algorithm performs better than both VQUPGMA and VQNJ (improvement of up to 25% in the quality of the clusters).

Rest of the paper is organized as follows: first, we give a brief overview of literature; second, we explain our VQSC algorithm; third, we present the numerical experiments; and fourth, we give conclusions and discuss future work.

# Literature Review

Clustering and sampling are very important techniques of machine learning, and these have been exhaustively researched. Thus, huge amount of information is available on these subjects. Hence, here we do not attempt to give a review of works done in these fields, rather we only present literature regarding usage of SC and sampling techniques in the field of plant genome.

SC can be performed in two ways; recursive and non-recursive. Bouaziz M. et al.[8] in 2012 used this method in a recursive way for genetic studies. However, we use a common non-recursive way[2, 9] since it is simpler and cheaper. It also gives tight and compact clusters.

As mentioned in the previous section, construction of the similarity matrix is the most important part of the SC algorithm. This can be done either by using basic techniques like Cosine Similarity, Alignment Score, Jukes Cantor, Pairwise Distance etc. or by using advanced techniques like Identity-by-State, Allele Sharing Distance, SNP edit distance, Covariance, Normalized Covariance, Coancestry etc.

Li L. et al.[10] in 2010 used SC for clustering gene sequences (which are a subset of WGSs) where they constructed the similarity matrix by Cosine Similarity. We use other basic techniques like Alignment Score, Jukes Cantor and Pairwise Distance as these capture the similarity between the SNP sequences in a better way. Lawson D. J. et al.[11] in 2012 used advance techniques of constructing the similarity matrix as mentioned above. We plan to use these as part of our future work.

Friedrich A. et al.[4] in 2007 used the strip method, the random method, the mean method, and the second derivative method to sample protein sequences (which are derived from gene sequences). However, we cannot use these as they use the E-values of the protein sequences which we do not have for our SNP sequences.

Zhang J. et al.[12] in 2011 used VQ to reduce the number of genome sequences of influenza A virus for better visualization of phylogenetic trees, which are an essential step in earlier mentioned clustering algorithms of UPGMA and NJ. They used the neural gas method as the basis of their sampling.

We use VQ as well but in a different sense. We use $k$-medoids as the basis of our sampling instead of the neural gas method. This is because, as earlier, it is easy to find the medoids of the kind of data we have (strings of A, T, G, C characters).

# The Vector Quantized Spectral Clustering (VQSC) Algorithm

The SC algorithm uses the concept of similarity graph to construct the similarity matrix (or the weighted adjacency matrix) that in turn is used to construct the Laplacian matrix (either normalized or non-normalized)[9]. Then, the eigenvectors corresponding to the first $k$ smallest eigenvalues (where $k$ is the number of clusters to be formed) of the Laplacian matrix are used to cluster the data.

As mentioned earlier, construction of the similarity matrix is significant in this algorithm because better the quality of this matrix, better is the accuracy of the SC algorithm. The Laplacian matrix obtained from the above-mentioned similarity matrix is also important because the eigenvectors of this matrix are used for clustering. A detailed description of the similarity matrix, the Laplacian matrix and the SC algorithm is given by Ulrike von Luxburg[9], Binkiewicz et al.[13], and Arias-Castro et al.[14].

In this paper, for constructing the similarity matrix, we compare every character in one SNP sequence with every character in other SNP sequences. This represents how much one sequence is different from another sequence. The dissimilarity $D(i,j)$ between any two SNP sequences $X_i$ and $X_j$ is defined as the number of positions at which $X_i$ and $X_j$ differ. The similarity value is calculated as

$$S(i,j) = l(seq) - D(i,j), \tag{1}$$

where $l(seq)$ is length of the SNP sequence. This value is normalized and used as the similarity value for $(i,j)$ index. We also use other similarity measures like Pairwise Distance[15], Jukes Cantor[16] and Alignment score[17] to construct the similarity matrix. Results show that the quality of clusters is sensitive to the quality of the similarity matrix used.

By nature of the algorithm, SC produces high-quality clusters on small-scale data, however as mentioned earlier, it has limited applicability to large-scale data because of its high computational complexity. Hence, we use VQ to compress the original data into a small set of representative data entities. The goal now is to minimize the difference between the original and this representative set.

Although the original VQ algorithm uses $k$-means, we achieve this minimized difference by using the $k$-medoids algorithm. This is because, as discussed earlier, data here is in the form of sequences of strings of A, T, G, C characters and mean of this data does not exist. On the other hand, $k$-medoids provides us with representative sequences from the set of given sequences itself.

Following is the algorithm for our VQSC-
Input: $n$ SNP sequences $\{x_i\}$ for $i = 1, \dots, n$; $k$ number of representative sequences to be selected; and $m$ number of clusters to be formed.
Output: clustered SNP sequences.

1. Perform $k$-medoids as follows:
    a. Compute medoids $y_1, \dots, y_k$ as the $k$ representative sequences.
    b. Build a correspondence table to associate each $x_i$ with the nearest medoid $y_j$.
2. Run the SC algorithm on $y_1, \dots, y_k$ to obtain cluster indexes $C_l$; $l = 1, \dots, m$ for each of $y_j$.

3. Recover the cluster membership for each $x_i$ by looking up the correspondence table.

# Numerical Experiments

We use SNP data of 31 Soybean sequences, which is taken from the database as follows:[18] http://chibba.pgml.uga.edu/snphylo/. This data contains 6,289,747 SNPs. Since this is a raw data, and hence, we use SNPhylo software[18] that removes low-quality data. Specifically, false SNPs are removed and we get 31 SNP sequences each of length 4847[‡]. Please refer to Figure 1 of Lee TH et al.[18], which shows the flowchart of SNPhylo pipeline (a commonly used standard procedure during preprocessing). Finally, these sequences are used to obtain the similarities amongst each other leading to the construction of the similarity matrix, which is an input to our VQSC algorithm.

Here, we first discuss the computational complexity of our and other standard algorithms (for SNP clustering). Next, we describe the criteria used to check the goodness of generated clusters, termed as validation metrics. Finally, we give our results.

## Computational Complexity

Table 1 given below compares the computational complexity of our VQSC algorithm with standard SC as well as earlier discussed and commonly used methods of UPGMA and NJ. Here, $n$ is the size of the data, $k$ is the number of representative samples for VQ, and $t$ is the number of iterations taken by VQ. Thus, it is evident from this table that use of VQ has made the SC algorithm computationally efficient ($n^2kt$ and $k^3$ are the computational complexities of VQ and the SC algorithm, respectively). As earlier, we are also better than UPGMA and NJ.

| Methods | Complexity | Comment |
|---|---|---|
| Spectral Clustering | $O(n^3)$ | Computationally less efficient, very good accuracy |
| **Our Algorithm (VQSC)** | $O(n^2kt + k^3)$ | **Computationally efficient, good accuracy** |
| UPGMA | $O(n^3)$ | Computationally less efficient, less accuracy |
| Neighbour Joining | $O(n^3)$ | Computationally less efficient, poor accuracy |

**Table 1:** Complexity analysis of different clustering algorithms.

Since, VQ reduces computational complexity of the underlying clustering algorithm substantially, in the results section we do two sets of comparisons; SC with UPGMA & NJ and VQSC with VQUPGMA & VQNJ.

## Validation Metrics

There are various metrics available for validation of clustering algorithms. These include Cluster Accuracy (CA)[1], Normalized Mutual Information (NMI)[1], Adjusted Rand Index (ARI)[1], Compactness (CP)[1], Separation (SP)[1], Davis-Bouldin Index (DB)[1], Silhouette Value[19] etc. For using the first three metrics, we should have prior knowledge of cluster labels. However, here we do not have ideal clustering results. Hence, we cannot use any of these validation metrics. Rest of the techniques do not have this requirement and hence, can be used for validation. We use Silhouette Value, which is usually used for validation of genome data[20].

‡ This software also constructs a phylogenetic tree as used by other standard genome clustering algorithms.

Silhouette Value is a measure of how similar an object is to its own cluster (intra-cluster similarity) compared with other clusters (inter-cluster similarity)[19]. For any cluster $C_l$ $(l = 1\ to\ m; say\ l1)$, let $a(i)$ be the average distance between the $i^{th}$ data point and all other points in cluster $C_{l1}$, and let $b(i)$ be the average distance between the $i^{th}$ data point and all other points in clusters $C_l$ $(l = 1\ to\ \ m\ and\ l\ != l1)$. Thus, Silhouette Value is given as

$$s(i) = \frac{b(i) - a(i)}{\max\{a(i),\ b(i)\}}\ .$$

Here, $a(i)$ and $b(i)$ signify the intra-cluster and the inter-cluster similarities, respectively. Silhouette Value lies between -1 and 1. A positive value indicates that the clusters are well-separated from each other, and a negative value indicates that the clusters are overlapping.

## Results

We first present the results of SC, UPGMA and NJ without VQ. This data is given in Table 2. Columns 2 to 5 refer to the Silhouette Values of the SC algorithm with four different similarity measures discussed earlier. Columns 6 and 7 give the Silhouette Values for UPGMA and NJ. As evident (highlighted in bold), SC with Alignment Score gives the best results for all the clusters.

| **# of Clusters** | **SC** | | | | **UPGMA** | **NJ** |
|---|---|---|---|---|---|---|
| | **Our Similarity *S(i, j)* from (1)** | **Pairwise Distance[15]** | **Jukes Cantor[16]** | **Alignment Score[17]** | | |
| 2 | 0.2012 | 0.2012 | 0.2590 | **0.3169** | 0.1831 | 0.2206 |
| 3 | 0.1987 | 0.1722 | 0.2440 | **0.2845** | 0.2002 | 0.2258 |
| 4 | 0.2053 | 0.2037 | 0.2621 | **0.3241** | 0.2546 | 0.2192 |
| 5 | 0.2488 | 0.2421 | 0.3017 | **0.3528** | 0.2791 | 0.2488 |
| 6 | 0.2771 | 0.2771 | 0.3214 | **0.3886** | 0.2389 | 0.2771 |
| 7 | 0.2990 | 0.3231 | 0.3414 | **0.3882** | 0.2612 | 0.2736 |
| 8 | 0.3451 | 0.3451 | 0.3811 | **0.4007** | 0.2906 | 0.2874 |
| 9 | 0.3490 | 0.3140 | 0.3785 | **0.4130** | 0.3112 | 0.3031 |
| 10 | 0.3522 | 0.3507 | 0.3771 | **0.4464** | 0.3430 | 0.2966 |
| 11 | 0.3687 | 0.3681 | 0.4045 | **0.4589** | 0.3831 | 0.3476 |
| 12 | 0.3799 | 0.4046 | 0.4258 | **0.5031** | 0.4089 | 0.3569 |
| 13 | 0.4329 | 0.3948 | 0.4611 | **0.5375** | 0.4153 | 0.3829 |
| 14 | 0.4470 | 0.4527 | 0.4646 | **0.5415** | 0.4610 | 0.4403 |

**Table 2:** Silhouette Values for different clustering algorithms without VQ.

Next, we discuss results for the same three clustering algorithms with VQ. VQ can be performed in two ways; either we can reduce the length of each sequence or we can reduce the number of sequences. In this work, we reduce the number of sequences to reduce complexity. Reducing the length of each sequence to reduce complexity is included in future work. Although with the decrease in the number of sequences, the accuracy of the clustering algorithm also decreases, however, in our case we show by experiments that this loss of accuracy is very less (acceptable when compared to the original non-VQ versions). Results for these experiments are given in Table 3 (structure of which is similar to that of Table 2). From this table, we see a similar pattern, i.e. our VQSC algorithm with Alignment Score is the best.

| # of Clusters | VQSC | | | | VQ UPGMA | VQNJ |
|---|---|---|---|---|---|---|
| | Our Similarity $S(i, j)$ from (1) | Pairwise Distance[15] | Jukes Cantor[16] | Alignment Score[17] | | |
| 2 | 0.2012 | 0.2012 | 0.2590 | **0.3169** | 0.1835 | 0.0128 |
| 3 | 0.2002 | 0.2002 | 0.2474 | **0.2876** | 0.2002 | 0.0427 |
| 4 | 0.2159 | 0.2181 | 0.2610 | **0.3052** | 0.2192 | 0.0752 |
| 5 | 0.2211 | 0.2488 | 0.2887 | **0.3232** | 0.2488 | 0.0827 |
| 6 | 0.2639 | 0.2528 | 0.2922 | **0.3046** | 0.2532 | 0.0476 |
| 7 | 0.2446 | 0.2184 | 0.2867 | **0.3259** | 0.2604 | 0.0821 |
| 8 | 0.2727 | 0.2718 | 0.3189 | **0.2935** | 0.2752 | 0.1195 |
| 9 | 0.2861 | 0.3209 | 0.2890 | **0.4004** | 0.2886 | 0.1506 |
| 10 | 0.3361 | 0.2429 | 0.3561 | **0.3726** | 0.3264 | 0.1523 |
| 11 | 0.3035 | 0.2877 | 0.3672 | **0.4594** | 0.3456 | 0.2273 |
| 12 | 0.3299 | 0.3783 | 0.4078 | **0.4743** | 0.3650 | 0.2513 |
| 13 | 0.4268 | 0.4184 | 0.3811 | **0.4843** | 0.4216 | 0.3002 |
| 14 | 0.4128 | 0.4251 | 0.4450 | **0.4966** | 0.4111 | 0.3465 |

**Table 3:** Silhouette Values for different clustering algorithms with VQ.

Finally, if we compare our best (VQSC with Alignment Score; Table 3 third last column) with existing best (UPGMA without VQ; Table 2 second last column), we are still about 20-25% better. As earlier, we also have the added benefit of reduced computational complexity.

Next, we further validate the quality of these clusters using tools used by biologists at Indian Institute of Soybean Research. Here, we compare cluster formation for SC and VQSC for the different number of clusters, i.e. $k$ = 11 and $k$ = 12 (see Figure 1 and Figure 2, respectively). We do not compare with UPGMA, VQUPGMA, NJ and VQNJ because the clusters obtained by SC and VQSC are of better quality. In these figures, the x-axis lists the 31 sequences and the y-axis refers to the clustering algorithms used. We use the data corresponding to Alignment Score as the similarity measure since that gives the best results. The Silhouette Values from Tables 2 and 3 are given on the right. The different colours denote the different clusters, and the coloured boxes signify which cluster each sequence belongs to.

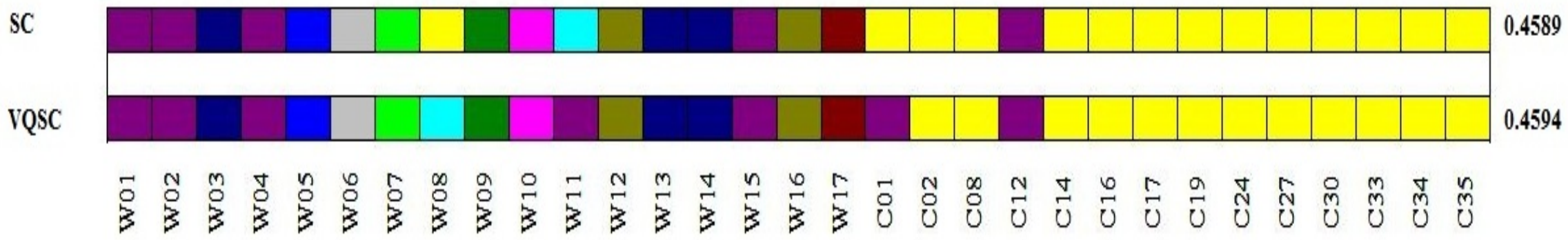

**Figure 1:** Cluster formation for SC and VQSC with Alignment Score and $k$ = 11.

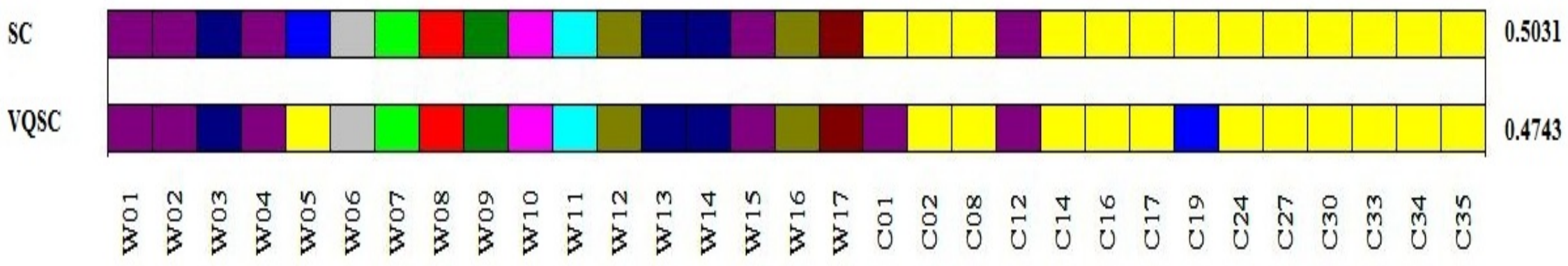

**Figure 2:** Cluster formation for SC and VQSC with Alignment Score and $k = 12$.

From Figure 1, we observe that our VQSC algorithm does not cluster sequences W08, W11 and C01 (i.e. only 3 out of 31) in their respective clusters when compared with SC. Similar behaviour can be observed from Figure 2. Sequences W05, C01, C19 (again only 3 out of 31) are not correctly clustered by VQSC when compared with SC. Thus, by using VQ with SC, we get almost same cluster formation but at a reduced computational cost.

# Conclusions and Future Work

We present the Vector Quantized Spectral Clustering (VQSC) algorithm that is a combination of Spectral Clustering (SC) and Vector Quantization (VQ) sampling. We use SC for its accurate clustering and VQ for its accurate sample selection. Use of this combination makes our algorithm scalable for large data as well. Since building the similarity matrix is critical to the SC algorithm, we exhaustively adapt four ways to build such a matrix for Soybean genome data. Adapting VQ for this data requires using $k$-medoids instead of traditional $k$-means for finding representative samples.

We compare the performance of our algorithms (SC and VQSC) with other traditional techniques (UPGMA, VQUPGMA, NJ and VQNJ). One variant of VQSC outperforms all other techniques both in-terms of cluster quality (improvement up to 25%) as well as computational complexity (order of magnitude faster).

In the future, we plan to extend this work to more number of sequences[21]. As earlier, here we reduce the number of sequences by sampling. However, we could also sample across the length of every sequence. Since the quality of the similarity matrix has a big impact on the quality of clusters, we also intend to adapt other ways of constructing this matrix as part of our future work[11].